\documentclass[page-classic,linenumbers]{epl2} 

\usepackage{graphics}
\usepackage{eurosym}
\usepackage{amsfonts}
\usepackage{amssymb}

\usepackage{amsmath}
\usepackage{graphicx}
\usepackage{xcolor}

\usepackage{amsmath,array}

\title{Hawking Radiation and Deflection of Light from Rindler Modified
Schwarzschild Black Hole}
\shorttitle{Rindler Modified Schwarzschild
Black Hole} 

\author{I. Sakalli \inst{1} \and A. Ovgun \inst{2,1}}
\shortauthor{Sakalli \etal}

\institute{                    
  \inst{1} Physics Department, Eastern Mediterranean University, Famagusta, Northern
Cyprus, Turkey\\
  \inst{2}  Instituto de F\'{\i}sica, Pontificia Universidad Cat\'olica de
Valpara\'{\i}so, Casilla 4950, Valpara\'{\i}so, Chile
}

\abstract{
We investigate the Hawking radiation of massive spin-1 vector particles, which are
coupled to vacuum fluctuations of a quantum field, from Rindler modified
Schwarzschild black hole. Rindler acceleration is used to produce the
post-general relativistic theory of gravity for the distant field of a point
mass. The gravitational lensing problem of the Rindler modified Schwarzschild
black hole is also studied. We compute the deflection angle for the IR region
(large distance limit as \textit{infrared}) by using the Gaussian curvature of
the optical metric of this back hole. Our investigations clarify how the
Rindler acceleration plays a role on the Hawking radiation and
gravitational lensing.}

\begin{document}

\maketitle



\section{Introduction}

Until the mid-1970s, black holes (BHs) were believed to be such super absorbent objects in the universe that nothing can come out from them. In 1974, Hawking \cite{SH1,SH2} showed that a BH can thermally create and emit virtual particles until it snuffs out its energy. This condition means that if a BH does not gain mass from any other source, then it could eventually shrink itself out of existence and evaporate completely; this process is called Hawking radiation (HR). HR is an intriguing puzzle that results from the amalgamation of general relativity and quantum mechanics. Hawking \cite{SH1,SH2} showed that black hole evaporation cannot be the result of unitary evolution of a pure state. Thus, the process of gravitational collapse is incompatible with the standard principles of quantum mechanics. In fact, the radiation emitted by a BH at late times is exactly described by a thermal density matrix (modulated by a greybody factor). A density matrix is a well-known matrix that describes a quantum system in a mixed state (a statistical ensemble of several quantum states). Therefore, as is the case with black-body radiation, HR quanta do not carry information. No mathematical transformation exists between the unitary operator and the density matrix. Thus, the information encoded in the wave function is irretrievably lost; this process is known as the so-called information loss paradox \cite{ilpR}. Efforts to resolve this problem have been continuing even today. A reader who wants to learn the details and recent developments about the information loss paradox may refer to \cite{ilpf,SH3,Sak1}. \textcolor{red}{On the other
hand, studies \cite{dense1,dense2,dense3} on the density matrix of a BH radiation have recently attracted much attention. It has been shown
that small fluctuations of the BH horizon give rise to corrections
to the density matrix of the HR. The associated corrections are shown to be
the correlations between the HR and the background geometry \cite{dense3}.
It is worth noting that correlations account for the information stored in
the BH with the collapsing matter.}
 Since the original studies of BH emission \cite{SH1,SH2}, 
\textcolor{red}{works on the HR have been still continuing. To date, HR was
verified by} various methods for many different types of particles having
spin $s=0,1/2,1,3/2,2,...$ \cite%
{PWT,parik,ang1,ang2,siri,sing1,sing2,sing3,masji,Sak0,Sak2,ao2,Sak3,Sak4,Vanz,mann0,mann1,mann2,mann3}%
. HR studies include the lower and higher dimensional BHs, wormholes, and
black strings \cite{SM0,ao6,SM1,SM2,SM3,SM4,SM5,SM6}. Recent studies \cite%
{stein1,stein2,stein3} have claimed that HR has been observed in the
laboratory environment. Those experiments about HR were conducted by
Steinhauer, who used a sonic (or the so-called analogue) 
\textcolor{red}{BH
in an atomic Bose-Einstein condensate \cite{bec} so that a state, where gas
bosons are cooled to temperatures very close to absolute zero (that is, very
near $0$K or $-273.16\,^{\circ}\mathrm{C}$)}. Thus, the sonic BH could mimic
a real cosmic event horizon. At this stage, Steinhauer managed to observe the particles of sound (phonons) at the BH's event horizon and found that sound waves in the Bose-Einstein condensate obey Hawking's theory \cite{stein3}. However, many physicists are still cautious about these results. Other experiments are needed to support the experiment of Steinhauer.

In particle physics, spin-1 particles are called vector particles. The most
well-known massive spin-1 particles, which are described by a three-dimensional spin
vector are the weak intermediate vector bosons $Z$ and $W^{\pm}$, and the
mesons $\boldsymbol{\Upsilon},$ $\mathbf{J}$/$\Psi,$ $\boldsymbol{\phi},$
and $\boldsymbol{\rho}$ \cite{bookP}. Photons are the massless spin-1
particles, which can only be directed parallel or anti-parallel to their
direction of motion. Free massive spin-1 fields are governed by the Proca
field equation, and massless spin-1 fields are governed by the Maxwell field equation.
Compelling evidence suggests that the spin-1 fields are potential dark
matter candidates \cite{DM1,DM2,DM3}. In recent years, HR of the spin-1
particles have attracted considerable interest \cite%
{VHR1,VHR2,VHR3,VHR4,VHR5,ao3,ao4}.

In this study, we consider the Rindler modified Schwarzchild BH\
(RMSBH) \cite{GBH1,GBH2}, which was initially proposed by Grumiller to
explain the mysterious attractive constant radial force that acts on the
Pioneer spacecrafts \cite{GBH3}. The force law obtained from Rindler's acceleration term has a considerable effect on the gravity at very long distances. However, Turyshev et al. \cite{Tury} showed
that the Pioneer anomaly is due to the thermal heat loss of the
satellites. Nevertheless, RMSBH is still on the agenda because it provides a basis for the theoretical explanation of the following issues: rotation curves of spiral galaxies, gravitational redshift, and perihelion shift in planetary orbits. Previous studies of the RMSBH, which include subjects of spectroscopy of area/entropy, quantum tunneling, and geodesics, can be seen in \cite{RM1,RM2,RM3}. \revision{The Rindler acceleration is also used in a quantum gravity corrected gravity theories to explain the rotation of the curve formula for the local galaxies \cite{GBH1,GBH2,GBH3}. The new Rindler acceleration term is studied as an alternative of the dark matter in galaxies and it is checked by using 8 galaxies of the HI Nearby Galaxy Survey which gives the Rindler acceleration parameter
of around $a \approx 3 * 10^{-9}$ $cm/s^2$ \cite{cin1,cin2,cin3,cin4}.}

\textcolor{red}{Light is electromagnetic radiation within a certain portion
of the electromagnetic spectrum. On the other hand, HR is mainly a thermal
radiation. The main purpose of this paper is to study the two possible problems
about the radiation physics (flow of atomic and subatomic particles and of
waves, such as those that characterize heat rays, light rays, and X-rays) of
the RMSBH: HR of the RMSBH and gravitational lensing of the RMSBH. To this
end, we first study the HR of the massive spin-1 particles tunneling from
the RMSBH. We apply the Hamilton-Jacobi
and the complex path integration quantum tunneling methods \cite{ang1,ang2,PAD} to the Proca equation \cite{VHR1} and obtain a set of differential equations.} Following \cite{VHR1,VHR2,VHR3,VHR4,VHR5}, we set the determinant of coefficient matrix of the equation to zero to obtain a non-trivial solution. Thus, we obtain the leading order term of the classical action ($S_{0}$ ) of the vector particles that are outgoing/ingoing from the horizon. We finally derive the tunneling rate of the spin-1 particles in the RMSBH and read the Hawking temperature of the RMSBH.
The phenomenon of gravitational lensing, which was predicted by Einstein's theory of general relativity, is a side effect of light moving along the curvature of spacetime, where the light that passes near a massive object is deflected slightly toward the mass. This phenomenon was observed for the first time in 1919 by Eddington and Dyson during a solar eclipse \cite{eclipse}. Since then, gravitational lensing has been one of the important tools in astronomy and astrophysics. For more details and recent contributions about gravitational lensing, the reader may refer to \cite{GL1,GL2,GL3,GL4,GL5,ne1,ne2}. We will also study the gravitational lensing problem of the RMSBH. For this purpose, we follow the geometrical method of Gibbons and Werner \cite{cqg}. In this manner, we explore the effect of the Rindler acceleration on the deflection of light moving in the IR region of the RMSBH.
The paper is organized as follows: In Section II, we introduce the physical features of the RMSBH geometry. Section III is devoted to the computation of the HR of the massive spin-1 particles from the RMSBH. In Section IV, we study the deflection of light from the RMSBH at the IR region via the method of Gibbons and Werner \cite{cqg}. Our results are summarized and discussed in Section V. \textcolor{red}{(Throughout the paper, we use units in which fundamental constants are $G=c=k_B= 1$)}.
\section{RMSBH Geometry}
Grumiller \cite{GBH1} constructed an effective model for the gravity of a central object at large scale \revision{(i.e. outside the galaxy)} called the RMSBH geometry. The Rindler term in the RMSBH spacetime causes an anomalous acceleration in the geodesics of test particles. RMSBH is the solution to the generic effective theory of gravity described by the following action:
\begin{equation}
S=-\int d^{2}x\sqrt{-g}\left[ \Phi^{2}R+2(\partial\Phi)^{2}+8a\Phi
-6\Lambda\Phi^{2}+2\right] ,  \label{1}
\end{equation}
where $g=\det(g_{
\mu
\nu})$ is the determinant of the metric tensor, $\Phi$ denotes the
scalar field, $\Lambda$ is the cosmological constant, $R$ is the Ricci
scalar and $a$ stands for the Rindler acceleration. \revision{The Rindler  acceleration is for explaining about the Pioneer anomaly and
also the shape of galactic rotation curve \cite{GBH1,GBH2,GBH3}}. After applying the
variational principle to action (1) and solving the corresponding field
equations, one can obtain the following spherically symmetric line-element
that models the IR\ gravity:
\begin{equation}
ds^{2}=-fdt^{2}+\frac{1}{f}dr^{2}+r^{2}\left( d\theta ^{2}+\sin ^{2}\theta
d\phi ^{2}\right) ,  \label{2}
\end{equation}
where 
\begin{equation}
f=1-\frac{2M}{r}-\Lambda r^{2}+2ar.  \label{3}
\end{equation}
Metric (2) is nothing but the RMSBH spacetime. In Eq. (3), the quantity $M$
is an integral constant. When $a=\Lambda =0$, one easily recovers the
Schwarzschild solution, with $M$ being the BH mass. Moreover, if $M=\Lambda
=0$, then line-element (2) is the two-dimensional Rindler metric \cite{RM}. The value of $\Lambda $ is approximated to $10^{-123}$ \cite{CMC1,CMC2}; therefore, we
set it to zero ($\Lambda =0$) for simplicity. Moreover, at the IR
region, the value of the Rindler acceleration is estimated as $a\approx
10^{-62}-10^{-61}$ \cite{GBH1}. 
Metric function (3) can be rewritten as 
\begin{equation}
f=\frac{2a}{r}(r-r_{h})(r-r_{n}),  \label{4}
\end{equation}
in which 
\begin{equation}
r_{n}=-\frac{1+\sqrt{1+16aM}}{4a}.  \label{5}
\end{equation}
\textcolor{red}{Since the radial coordinate $r$ is defined in the range of
$0\leq r<\infty$, it is therefore clear that $r_{n}$ (having negative value)
is not a physical quantity, and thus it cannot be interpreted as the horizon
(see for example \cite{Cataldo}). So, RMSBH spacetime possesses only one
horizon, which is called the event horizon ($r_{h}$):} 
\begin{equation}
r_{h}=\frac{-1+\sqrt{1+16aM}}{4a}.  \label{6}
\end{equation}
The Bekenstein-Hawking entropy \cite{RM,Wbk} of the RMSBH is given by 
\begin{equation}
S_{BH}=\frac{A_{h}}{4\hbar}=\frac{\pi r_{h}^{2}}{\hbar},  \label{7}
\end{equation}
where $A_{h}=4\pi r_{h}^{2}$ is the surface area of the RMSBH. The following
conventional definition of the surface gravity defined for a spherically
static spacetime \cite{Wbk} is used:
\begin{equation}
\kappa =\left. \frac{\partial _{r}f}{2}\right\vert _{r=r_{h}}=a\left( 1-%
\frac{r_{n}}{r_{h}}\right) ,  \label{8}
\end{equation}
where a prime on a function denotes differentiation with respect to $r$.
From here on, we read the Hawking temperature of the RMSBH as follows:
\begin{equation}
T_{H}=\frac{\hbar\kappa}{2\pi}=\frac{a\hbar}{2\pi}\left( 1-\frac{r_{n}}{r_{h}%
}\right) .  \label{9}
\end{equation}
Figure 1 shows how the Hawking temperature changes with Rindler acceleration
in the RMSBH geometry. 
\begin{figure}[h]
\centering
\includegraphics[width=8cm,height=6cm]{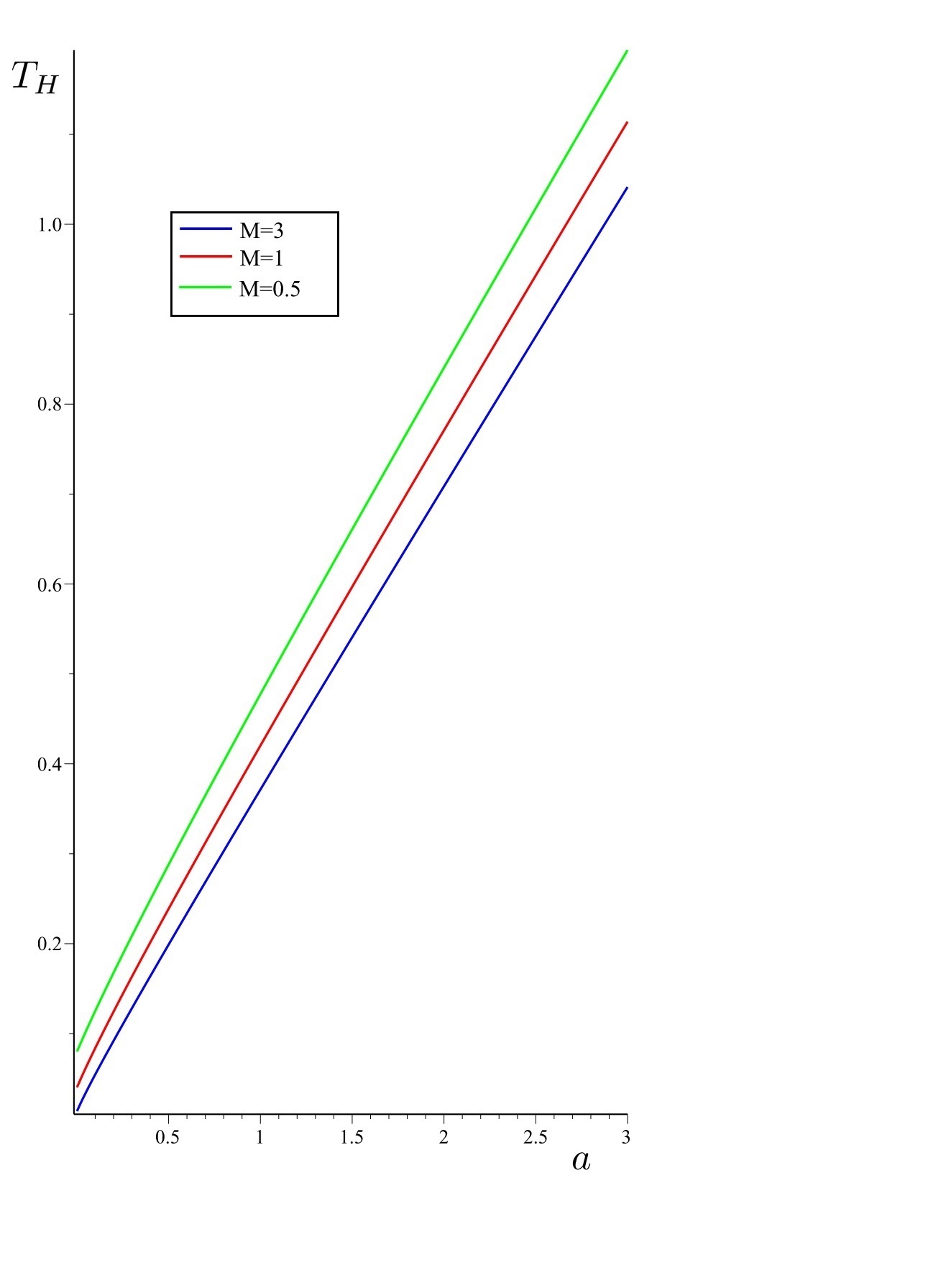}
\caption{Plots of $T_{H}$ versus $a$. The plots are governed by Eq. (9).
Each color represents a RMSBH with different mass.}
\end{figure}
\section{Quantum Tunneling of Massive Vector Particles from RMSBH}
In this section, we compute the HR of massive vector particles, which
quantum mechanically tunnel from the RMSBH. We consider the Proca equation 
\cite{VHR1}:
\begin{equation}
\frac{1}{\sqrt{-g}}\partial_{\mu}\left( \sqrt{-g}\Psi^{\nu\mu}\right) +\frac{%
m^{2}}{\hbar^{2}}\Psi^{\nu}=0,  \label{10}
\end{equation}
which corresponds to the wave equation of a spin-1 field $\Psi_{\nu}$ of
mass $m.$ In Eq. (10), the second rank tensor is defined by 
\begin{equation}
\Psi_{\mu\nu}=\partial_{\mu}\Psi_{\nu}-\partial_{\nu}\Psi_{\mu}.  \label{11}
\end{equation}
\textcolor{red}{Considering the WKB method, ansatz of the spin-1 field
can be defined by \cite{VHR2,Banar}}
\begin{equation}
\Psi_{\nu}=C_{\nu}\exp\left( \frac{i}{\hbar}\left( S_{0}(t,r,\theta
,\phi)+\hbar\,S_{1}(t,r,\theta,\phi)+\hdots.\right) \right) ,  \label{12}
\end{equation}
\textcolor{red}{where $S_{0}(t,r,\theta,\phi)$  is the kinetic term equal to the classical action of the particles \cite{Vanz}, $S_{j=1,2,..}(t,r,\theta,\phi)$ are the higher order action corrections, and $C_{\nu}=(C_{1},C_{2},C_{3},C_{4})$ represents some arbitrary constants.
Furthermore, taking cognizance of the Hamilton-Jacobi method which takes the
advantage of the symmetry (Killing vectors) of the spacetime \cite{Banar}, one can set the leading order of
the action to \cite{Chen2015}} 
\begin{equation}
S_{0}(t,r,\theta ,\phi )=-Et+R(r,\theta )+j\phi +k,  \label{13}
\end{equation}%
where $E$\ and $j$ denote the energy and angular momentum of the vector
particles, respectively. $k$ is a complex constant. 
 
After manipulating the Proca equation (10) with Eq.s (12) and (13), we obtain a quadruple equation set to the lowest order in $\hbar $ as follows
\begin{multline}
-E\left( \partial _{r}R\right) C_{1}-\frac{E(\partial _{\theta }R)C_{2}}{%
r^{2}f}-\frac{\left[ \sin ^{2}\theta \left( r^{2}f(\partial
_{r}R)^{2}+m^{2}r^{2}+(\partial _{\theta }R)^{2}\right) +j^{2}\right] C_{4}}{%
r^{2}f\sin ^{2}\theta }
\\
-\frac{EjC_{3}}{r^{2}f\sin ^{2}\theta }=0,
\end{multline}
\begin{multline}
-\frac{\left[ -\sin ^{2}\theta f(\partial _{\theta }R)^{2}-f(m^{2}r^{2}\sin
^{2}\theta +j^{2})+E^{2}r^{2}\sin ^{2}\theta \right] C_{1}}{r^{2}\sin
^{2}\theta }-\frac{f\left( \partial _{r}R\right) \left( \partial _{\theta
}R\right) C_{2}}{r^{2}}
\\ 
-\frac{jf\left( \partial _{r}R\right) C_{3}}{r^{2}\sin \theta }-E\left(
\partial _{r}R\right) C_{4}=0,  \label{15}
\end{multline}
\begin{multline}
-\frac{f\left( \partial _{r}R\right) (\partial _{\theta }R)C_{1}}{r^{2}}-%
\frac{\left[ -\sin ^{2}\theta f\left( \partial _{r}R\right)
^{2}-f(m^{2}r^{2}\sin ^{2}\theta +j^{2})+E^{2}r^{2}\sin ^{2}\theta \right]
C_{2}}{r^{2}\sin ^{2}\theta }
\\
-\frac{j(\partial _{\theta }R)C_{3}}{r^{4}\sin ^{2}\theta }-\frac{(\partial
_{\theta }R)EC_{4}}{r^{2}f}=0,  \label{16}
\end{multline}
\begin{multline}
-\frac{jf\left( \partial _{r}R\right) C_{1}}{r^{2}\sin ^{2}\theta }-\frac{%
j(\partial _{\theta }R)C_{2}}{r^{4}\sin ^{2}\theta }+\frac{\left[ f(\partial
_{\theta }R)^{2}-r^{2}(-f^{2}\left( \partial _{r}R\right) ^{2}-m^{2}f+E)%
\right] C_{3}}{r^{4}f\sin ^{2}\theta }
\\
-\frac{jEC_{4}}{r^{2}\sin ^{2}\theta f}=0,  \label{17}
\end{multline}

After this stage, we can obtain a $4\times4$ coefficient matrix of $%
\aleph\left(C_{1},C_{2},C_{3},C_{4}\right)^{T}=0$ in which the superscript $%
T $ represents the transition to the transposed vector. The non-zero
components of $\aleph$ matrix are given by
\begin{align}
\aleph _{11}& =\aleph _{24}=-E\left( \partial _{r}R\right) , \\
\aleph _{12}& =\aleph _{34}=-\frac{E(\partial _{\theta }R)}{r^{2}f}, \\
\aleph _{13}& =\aleph _{44}=-\frac{Ej}{r^{2}f\sin ^{2}\theta }, \\
\aleph _{14}& =-\frac{\left[ \sin ^{2}\theta \left( r^{2}f\left( \partial
_{r}R\right) ^{2}+m^{2}r^{2}+(\partial _{\theta }R)^{2}\right) +j^{2}\right] 
}{r^{2}f\sin ^{2}\theta }, \\
\aleph _{21}& =-\frac{\left[ -\sin ^{2}\theta f(\partial _{\theta
}R)^{2}-f(m^{2}r^{2}\sin ^{2}\theta +j^{2})+E^{2}r^{2}\sin ^{2}\theta \right]
}{r^{2}\sin ^{2}\theta }, \\
\aleph _{22}& =\aleph _{31}=-\frac{f\left( \partial _{r}R\right) (\partial
_{\theta }R)}{r^{2}},
\end{align}
\begin{align}
\aleph _{23}& =\aleph _{41}=-\frac{jf\left( \partial _{r}R\right) }{%
r^{2}\sin \theta }, \\
\aleph _{32}& =-\frac{\left[ -\sin ^{2}\theta f\left( \partial _{r}R\right)
^{2}-f(m^{2}r^{2}\sin ^{2}\theta +j^{2})+E^{2}r^{2}\sin ^{2}\theta \right] }{%
r^{2}\sin ^{2}\theta }, \\
\aleph _{33}& =\aleph _{42}=-\frac{j(\partial _{\theta }R)}{r^{4}\sin
^{2}\theta }, \\
\aleph _{43}& =\frac{\left[ f(\partial _{\theta }R)^{2}-r^{2}(-f^{2}\left(
\partial _{r}R\right) ^{2}-m^{2}f+E)\right] }{r^{4}f\sin ^{2}\theta }.
\end{align}

The non-trivial solution of the quadruple equation set is conditional on $%
\det \aleph =0$. Thus, we have 

\begin{equation}
\frac{m^{2}\left[ \sin ^{2}\theta f(\partial _{\theta
}R)^{2}+f^{2}r^{2}\left( \partial _{r}R\right) ^{2}+f\left( m^{2}\sin
^{2}\theta r^{2}+j^{2}\right) -E^{2}r^{2}\sin \theta \right] ^{3}}{%
r^{10}\sin \theta f^{3}}=0,
\end{equation}

which yields the integral solution of the radial function as
\begin{equation}
R_{\pm}=\int\pm\frac{\sqrt{E^{2}-f\left( m^{2}+\frac{(\partial_{\theta}R)^{2}%
}{r^{2}}+\frac{j^{2}}{\sin^{2}\theta r^{2}}\right) }}{f}.  \label{29n}
\end{equation}

$R_{+}(r)$ and $R_{-}(r)$ correspond to the outgoing (i.e., emission) and
ingoing (i.e., absorption) vector particles from/to the RMSBH, respectively.
Because of a pole located at the horizon, the imaginary part of $%
R_{\pm}(r)$ can be calculated by using the complex path integration method 
\cite{ang1,ang2}. Therefore, we can evaluate the integral (29) in the
vicinity of the horizon as
\begin{equation}
ImW_{\pm }(r)=\pm \left. \frac{\pi }{\partial _{r}f}E\right\vert _{r=r_{h}}.
\label{30}
\end{equation}%

The probabilities of the massive vector particles, tunneling from the
horizon of RMSBH, become
\begin{align}
P_{emission} & =e^{-\frac{2}{\hbar}ImS_{+}}=e^{\left[ -\frac{2}{\hbar }%
(ImW_{+}+Im\Bbbk)\right] },  \label{32} \\
P_{absorption} & =e^{-\frac{2}{\hbar}ImS_{-}}=e^{\left[ -\frac{2}{\hbar }%
(ImW_{-}+Im\Bbbk)\right] }.  \label{27}
\end{align}
According to the classical concept of BH physics, the ingoing vector particles
must be fully absorbed, as indicated by $P_{absorption}=1$. This requirement can be fulfilled with $Im\Bbbk=-ImW_{-}$. Recalling $W_{+}=-W_{-}$, we can thus
compute the quantum tunneling rate of the massive vector particles of the
RMSBH as
\begin{align}
\Gamma =P_{emission}=\exp \left( -\frac{4}{\hbar }ImW_{+}\right) \notag \\ 
=\exp
\left( -\left. \frac{4\pi }{\hbar \left( \partial _{r}f\right) }E\right\vert
_{r=r_{h}}\right) .  \label{28}
\end{align}%
$\Gamma $ is also equivalent to the Boltzmann factor \cite{Vanz}. The latter
remark enables us to compute the surface temperature of the RMSBH as follows:
\begin{equation}
T=\left. \frac{\hbar \left( \partial _{r}f\right) }{4\pi }\right\vert
_{r=r_{h}}=\frac{a\hbar (r_{h}-r_{n})}{2\pi r_{h}}.
\end{equation}
The above result completely overlaps with the Hawking temperature (9) of the
RMSBH. The Hawking temperature of the RMSBH increases proportionally with the Rindler acceleration, which can be also best seen in Fig. (1). 
\section{Deflection of Light from RMSBH}
In this section, we will study the deflection of null geodesics from the RMSBH, that is, the gravitational lensing problem.
\revision{ The main reason to combine the Hawking radiation and the deflection of light is that the gravitational lensing is a observational effect to help to model the theoretical theories, made new claim that due to the interaction of the galaxies, the normal matter is separated from the dark matter after collision of galaxies then the bending of light is occurred due to the the dark matter from the cluster differently to the case without dark matter.  Moreover, increasing number of new observations give evidences to propose that the  unknown particles of dark matter may have another property that is that they can make self-interaction, namely an exchange of dark photons which have spin-1 may create the force. In this work, by investigating radiation of spin-1 particles, such as dark photons from the black hole, we find the evidence of dark matter. Moreover, astronomers may have an extremely powerful tool to seeing the gravitational effects of the dark matter using the gravitational lensing. On this purpose, weak lensing occurs when the light from a distant galaxy passes a distance from a dark-matter concentration and produces a slight distortion in the shape of a distant galaxy.}

 To this end, we will suppose that RMSBH consists of a perfect fluid, which can be thought of as the stellar fluid of a cluster of galaxies. Such stellar fluids act as a gravitational lens, which is capable of bending the light from the source as the light travels toward the observer.
By using the static and spherically symmetric feature of metric (2), without loss of generality, we can assume that null geodesics ($ds^{2}=0$) lie in the equatorial plane: $\theta=\pi/2$. Therefore, all images are collinear with the lens center \cite{LC}. Light rays are the spatial projection of the null geodesics of the line element of the optical metric \cite{OM}, which is given by
\begin{equation}
dt^{2}=g_{ij}^{om}dx^{i}dx^{j},  \label{35}
\end{equation}
where $g_{ij}^{om}$ originates from Fermat's principle {\cite{FP,FP1}}%
:
\begin{equation}
g_{ij}^{om}=\frac{g_{ij}}{-g_{00}}.  \label{36}
\end{equation}
Thus, metric (35) becomes
\begin{equation}
dt^{2}=dr^{\ast2}+F(r^{\ast})^{2}d\phi^{2},  \label{37}
\end{equation}
where $r^{\ast}$ is the radial Regge-Wheeler tortoise coordinate
\begin{equation}
dr^{\ast }=\frac{dr}{f},  \label{38}
\end{equation}
and
\begin{equation}
F(r^{\ast })=\frac{r}{\sqrt{f}},\text{ \ \ \ \ }r=r(r^{\ast }).  \label{39}
\end{equation}
We immediately deduce from Eq. (37) that the optical metric is a surface of
revolution. The intrinsic or the so-called Gaussian curvature $\mathcal{K}$ {%
\cite{cqg}} of the optical metric can be expressed as follows:
\begin{align}
\mathcal{K} & \mathcal{=}-\frac{1}{F(r^{\ast})}\frac{d^{2}F(r^{\ast})}{%
dr^{\ast2}},  \notag \\
& =-\frac{1}{F(r^{\ast})}\left[ \frac{dr}{dr^{\ast}}\frac{d}{dr}\left( \frac{%
dr}{dr^{\ast}}\right) \frac{dF(r^{\ast})}{dr}+\left( \frac {dr}{dr^{\ast}}%
\right) ^{2}\frac{d^{2}F(r^{\ast})}{dr^{2}}\right] .  \label{40}
\end{align}
The area element of the optical metric (35) is given by
\begin{equation}
dA=\sqrt{\left\vert \det g^{0m}\right\vert }drd\phi =\frac{r}{\sqrt{f^{3}}}%
drd\phi .  \label{41}
\end{equation}
We now consider a lens model of perfect fluid, which is characterized by the
following energy momentum tensor {\cite{XXX}}:
\begin{equation}
T^{\alpha\beta}=diag[\rho(r),p(r),p(r),p(r)],  \label{42}
\end{equation}
where $p(r)$\ and $\rho (r)$ denote the pressure and density, respectively. %
\textcolor{red}{If we set \cite{XXX}}
\begin{equation}
f=\frac{r-2\mu (r)}{r},  \label{43}
\end{equation}
\textcolor{red}{where}
\begin{equation}
\mu (r)=4\pi \int\limits_{0}^{r}\rho (r^{\prime })r^{^{\prime }2}dr^{\prime
},  \label{44n}
\end{equation}
\textcolor{red}{from Eq.s (3), (43), and (44), we obtain}
\begin{equation}
\int\limits_{0}^{r}\rho (r^{\prime })r^{^{\prime }2}dr^{\prime }=\frac{1}{%
4\pi }\left( M-ar^{2}\right) .  \label{45n}
\end{equation}
Therefore, the Tolman-Oppenheimer-Volkoff equation {\cite{TOV,XXX}}, which
ensues from the conservation of the energy momentum tensor ($\nabla _{\alpha
}T^{\alpha \beta }=0$), reads \cite{XXX}
\begin{equation}
\frac{dp(r)}{dr}=-\frac{\left[ \rho (r)+p(r)\right] \left[ \mu (r)+4\pi
r^{3}p(r)\right] }{r^{2}f}.  \label{46}
\end{equation}
Equation (46), in fact, shows the hydrodynamical behavior of a massive
compact astrophysical object in terms of mass, pressure, and energy
density \cite{XXX}. We also have
\begin{equation}
\frac{df}{dr}=\frac{2}{r^{2}}\left[ \mu (r)+4\pi r^{3}p(r)\right] .
\label{47}
\end{equation}
By inserting Eq.s (46) and (47) into Eq. (40), we can obtain the Gaussian optical
curvature as follows:
\begin{equation}
\mathcal{K=}\frac{-2\mu (r)}{r^{3}}\left\{ f+\frac{\mu }{2r}-\frac{4\pi r^{3}%
}{\mu }\left[ \rho (r)f+p(r)\left( f-\frac{\mu (r)}{r}-2\pi p(r)r^{2}\right) %
\right] \right\} .  \label{48}
\end{equation}
Our main motivation is to study the weak gravitational lensing phenomenon,
which is more realistic in the astrophysical observation. For this reason,
we shall focus on the non-relativistic kinetic theory (e.g., the
collisionless Boltzmann equation \cite{BE} of the non-relativistic stellar
fluids). This condition means that the pressure is ignored in Eq. (48) (see also \cite%
{XXX}). Thus, the multiplication of the Gaussian optical curvature with the
area element of the optic metric results in
\begin{equation}
\mathcal{K}dA=\frac{-2\mu (r)}{r^{2}\sqrt{f^{3}}}\left\{ f+\frac{\mu (r)}{2r}%
-\frac{4\pi r^{3}}{\mu (r)}\left[ \rho (r)f\right] \right\} drd\phi .
\label{49}
\end{equation}
\textcolor{red}{In the spirit of \cite{cqg}, RMSBH lens can be characterized by [see Eq.
(45)]}
\begin{equation}
\rho (r)\approx -\frac{a}{2\pi r}.  \label{50n}
\end{equation}
\textcolor{red}{However, in the weak field limit [corresponding to
$r>>2M$ (see for example \cite{wfl1,wfl2})], similar to the
Schwarzschild lens \cite{cqg}, the density vanishes. Therefore, Eq.
(44) admits}
\begin{equation}
\mu (r)\approx \text{constant},  \label{51n}
\end{equation}
\textcolor{red}{and Eq. (49) becomes}
\begin{equation}
\mathcal{K}dA\approx \frac{-2\mu (r)}{r^{2}\sqrt{f^{3}}}\left[ f+\frac{\mu
(r)}{2r}\right] drd\phi .  \label{52n}
\end{equation}
\textcolor{red}{The deflection angle ($\delta$) is calculated by integrating the Gaussian
curvature on the circular boundary of domain $D_{2}$ \cite{cqg}, which can be best seen from \textit{figure 1} of \cite{cqg}. In the
weak deflection limit, it is assumed that the light rays are governed by the
function (at zeroth order) of}
\begin{equation}
r(t)=\frac{b}{\sin \phi }  \label{53n}
\end{equation}
\textcolor{red}{where $b$ is known as the impact parameter with $b\gg M$
\cite{XXX}. Furthermore, since we take cognizance of the IR region, Eq. (52) can be recast into the following expression by using the Taylor
series}
\begin{equation}
\mathcal{K}dA\approx -\frac{1}{2\sqrt{2}}\left( ar^{3}\right) ^{-\frac{3}{2}%
}drd\phi .  \label{54n}
\end{equation}
Thus, $\delta$ is computed via the following integral {\cite{cqg}}
\begin{align}
\delta & =-\int \int_{D_{2}}\mathcal{K}dA,  \notag \\
& \approx \frac{1}{2\sqrt{2}}\int_{0}^{\pi }\int_{b/\sin \phi }^{\infty
}\left( ar^{3}\right) ^{-\frac{3}{2}}drd\phi ,  \notag \\
& =\frac{10}{147\sqrt{a^{3}b^{7}}}EllipticK\left( \frac{1}{\sqrt{2}}\right) ,
\notag \\
& \simeq \frac{0.126127529}{\sqrt{a^{3}b^{7}}}.  \label{55n}
\end{align}
The above result shows the RMSBH's gravitational lensing deflection angle at
the IR region. Evidently, the deflection angle is inversely proportional to the Rindler acceleration, that is, gravitational lensing of an accelerated RMSBH is less than the almost non-accelerated RMSBH. The latter remark implies that observing (via the gravitational lensing) RMSBH will be more difficult than observing Schwarzschild BH. Meanwhile, we should not overlook the point that $a^{3}b^{7}$ would have a significant value because the impact parameter $b\gg\mu$ in the denominator of Eq. (55). This finding means that the deflection angle possesses a small value at the IR region.
\section{Conclusion}
\textcolor{red}{In this paper, two problems of radiation physics about the RMSBH were considered. We focused on the HR of massive spin-1 particles and
gravitational lensing of the RMSBH, respectively. In fact, massive spin-1
particles are a potential dark matter candidate \cite{DM1} and one of the
most successful techniques to explore the dark matter has so far been the
effect of gravitational lensing \cite{Dmat2}.} \textcolor{red}{Although our work is purely theoretical, we believe that such theoretical studies will lead to the observation of the dark matter in future.}
To investigate the HR of massive vector particles, we employed the Proca equation (10) in the RMSBH background within the concept of semiclassical WKB approximation. By taking into account the Hamilton-Jacobi  and the complex path integration methods \cite{ang1,ang2}, we computed the tunneling rate (28) of the RMSBH, which is ruled by the well-known Boltzmann factor. Then, the temperature obtained from the resulting tunneling rate is shown to be same as the original Hawking temperature (9) of the RMSBH. The Hawking temperature of the RMSBH increases proportionally with the Rindler acceleration, which can be best seen in Fig. (1). We also applied the geometrical approach of the gravitational lensing theory to the RMSBH. The optical metric (37) in which the geodesics are the spatial light rays of the RMSBH was derived to calculate the Gaussian curvature in the weak deflection limit. At the IR region, deflection angle (54) decreases with the increasing Rindler acceleration value, that is, the gravitational lensing of the RMSBH is always less than the Schwarzschild BH. 
The gravitational lensing of a rotating RMSBH, which can be easily obtained by the Newman-Janis algorithm \cite{NJA,NJA2}, would be interesting to explore. In a rotating geometry, images are no longer collinear with the lens center \cite{GLI}. Therefore, such a problem may reveal more information compared with the present case. This issue is the next stage of study that interests us. 
\acknowledgements
This work was supported by the Chilean FONDECYT Grant No. 3170035 (A\"{O}).
\revision{We thank the Editor and anonymous Referee for their constructive comments and suggestions, which helped us improve the manuscript.}

\end{document}